\documentclass{article}

\usepackage{PRIMEarxiv}

\usepackage[utf8]{inputenc} 
\usepackage[T1]{fontenc}    
\usepackage{hyperref}       
\usepackage{url}            
\usepackage{booktabs}       
\usepackage{amsfonts}       
\usepackage{nicefrac}       
\usepackage{microtype}      
\usepackage{lipsum}
\usepackage{fancyhdr}       
\usepackage{graphicx}       
\usepackage{comment}
\graphicspath{{media/}}     

\title{Heuristic-Based Address Clustering in Cardano Blockchain
}
\author{
Mostafa Chegenizadeh\thanks{M. Chegenizadeh (mostafa.chegenizadeh@uzh.ch) is a PhD Candidate in the Blockchain \& Distributed Ledger Technologies Group at the Department of Informatics, University of Zurich, Switzerland.},
Sina Rafati Niya\thanks{S. Rafati Niya (sina.rafatiniya@uzh.ch) is a Senior Research Associate in the Blockchain \& Distributed Ledger Technologies Group at the Department of Informatics, University of Zurich, Switzerland.},
Claudio J. Tessone\thanks{C.J. Tessone (claudio.tessone@uzh.ch) is Professor of Blockchain and Distributed Ledger Technologies at the Department of Informatics, University of Zurich, Switzerland. He is also co-founder and Chairman of the UZH Blockchain Center.}
}

\begin{document}
\maketitle

\begin{abstract}
Blockchain technology has recently gained widespread popularity as a practical method of storing immutable data while preserving the privacy of users by anonymizing their real identities. This anonymization approach, however, significantly complicates the analysis of blockchain data. To address this problem, heuristic-based clustering algorithms as an effective way of linking all addresses controlled by the same entity have been presented in the literature. In this paper, considering the particular features of the Extended Unspent Transaction Outputs accounting model introduced by the Cardano blockchain, two new clustering heuristics are proposed for clustering the Cardano payment addresses. Applying these heuristics and employing the UnionFind algorithm, we efficiently cluster all the addresses that have appeared on the Cardano blockchain from September 2017 to January 2023, where each cluster represents a distinct entity. The results show that each medium-sized entity in the Cardano network owns and controls 9.67 payment addresses on average. The results also confirm that a power law distribution is fitted to the distribution of entity sizes recognized using our proposed heuristics.
\end{abstract}


\section{Introduction}
Blockchain technology has attracted much attention in recent years as a revolutionary way of immutable data storage that has enabled a new generation of financial exchange platforms. In contrast to traditional banking systems, blockchain technology proposes a promising decentralized method enabling users to transfer value/assets to other users without the need for the presence of any centralized trusted third party to act as an intermediary in the network. 
In a blockchain network, each node is connected to and exchanges data with an arbitrary number of desired nodes, forming a peer-to-peer network. 
The main purpose of this decentralized network is to reach an agreement on a unique ledger, which is an ordered chain of transactional data.
The key idea is that there is no central authority in a blockchain-based network.
The transactions are generated by the network users and sent to the peers. The peers validate the transactions, put them together in blocks, store the blocks in their local ledger, and propagate the blocks to other peers. In this way, the transactions are broadcasted throughout the whole network, ensuring that a common record of transactions is shared among all peers.

The basic decentralized consensus protocol introduced by Satoshi Nakamoto in 2008 was called Proof of Work (PoW)\cite{b2}. 
In PoW, each peer needs to complete a computationally heavy task before it can publish a block of transactions. 
The first peer that completes this task and publishes the block will be awarded according to the protocol. In fact, finding the appropriate parameters for a block is a time- and computation-intense task certifying some intensive work has been done by the peer, hence the name "Proof of Work". 
However, this heavy computation is used only to prove that some "work" is done, while the "work" itself is in some respects useless---at least avoidable---in the protocol. Thus, this "work" can be regarded as an inefficient utilization of time and energy. 
To address this problem, alternative consensus protocols have been introduced after PoW. 
One of the alternative protocols is called Proof of Stake (PoS)\cite{b19}. In PoS, only stake owners, or users who have a certain amount of cryptocurrency in their possession, are in authority to validate new blocks. The next block validator is selected randomly with a higher probability of selecting a stake owner with a larger stake. This setting means that the selected validator does not need to perform heavy computations, but only validates the block's transactions.

Blockchains are supposed to determine the provenance and ownership of assets in the network. To this end, the existing blockchains use two different accounting models\cite{b21}.
The first model is called Unspent Transaction Output (UTXO)-based model, which is similar to traditional cash trading. 
In this model, each transaction spends a finite number of inputs and generates some outputs with fixed values, where the inputs are unspent outputs of previous transactions, and the total amount of inputs must exceed the total amount of outputs. 
UTXOs are indeed transaction outputs that have not been spent yet. Each output is locked by a specific public address, which identifies a particular entity in the network allowed to spend it.
Since the value of each UTXO is determined in the transaction generating it, a UTXO owner who intends to spend it cannot divide it but instead must spend the entire amount and receive a change value in return.
Bitcoin\cite{b2}, which is a cryptocurrency with the largest market capacity\cite{b47}, exemplifies this UTXO-based accounting model. 
The second model is called the account-based model, which is closely similar to the banking system; where each user owns one or more accounts, and the transactions represent a transfer of an arbitrary amount of value from one account to another. 
In this setting, the blockchain can be considered as a state machine that records the current balances of all accounts, where each transaction changes the state of the blockchain by updating the balances: subtracting the input value from the sender's account and adding the output value to the receiver's account.
Account-based model is exemplified by Ethereum\cite{b16}, which is the second-largest cryptocurrency in the world\cite{b47}. Ethereum is also the first blockchain platform that presented the concept of smart contracts.

In contrast to the banking system, the details of all transactions are publicly available in public blockchains, which can threaten the privacy of users. For this reason, in many UTXO-based blockchains (e.g., Cardano, Bitcoin, Litecoin, and Monacoin), each user/entity has been enabled to generate and control an arbitrary number of anonymous addresses. Furthermore, many crypto wallets (e.g., Daedalus, Yoroi, and Coinbase), by default, generate new change addresses for the users/entities after each transaction.
Accordingly, it is reasonable for data analysts to assume that some blockchain users/entities are expected to own multiple addresses.\\

\begin{figure}[h]
    \centering
    \includegraphics[width=0.65\textwidth]{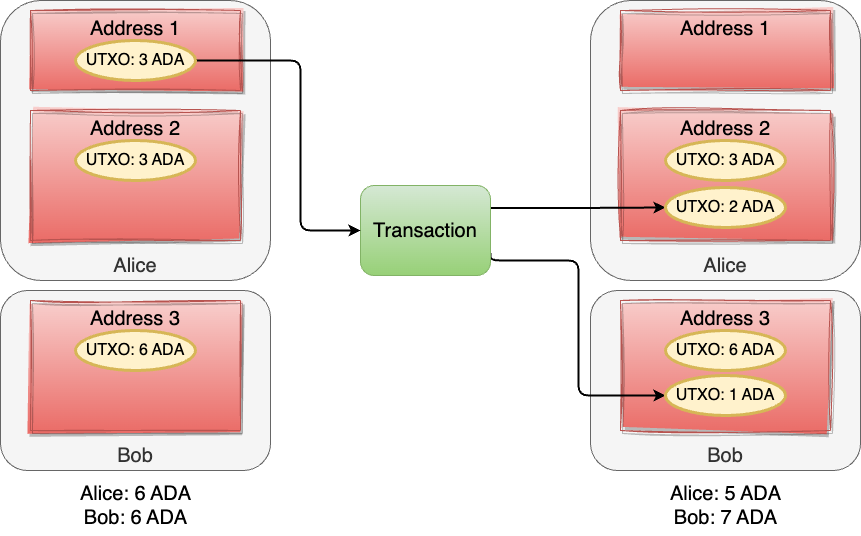}
    \caption{\small Example scenario 1: value transfer among different addresses of the same entity\\}
    \label{fig:scenario_1}
\end{figure}

Considering that it is necessary for many financial analyses---for example, wealth distribution or asset velocity---to take into account the transactions between distinct entities rather than distinct addresses, address clustering would be the first step to conducting a statistical analysis on a UTXO-based blockchain in many cases.
In other words, to analyze the network parameters of a UTXO-based blockchain, data analysts cannot treat each address as if it belongs to a distinct entity; instead, they need to cluster all addresses controlled by the same entity together and then perform their analyses based on those clusters.\\

\begin{figure}[h]
    \centering
    \includegraphics[width=0.65\textwidth]{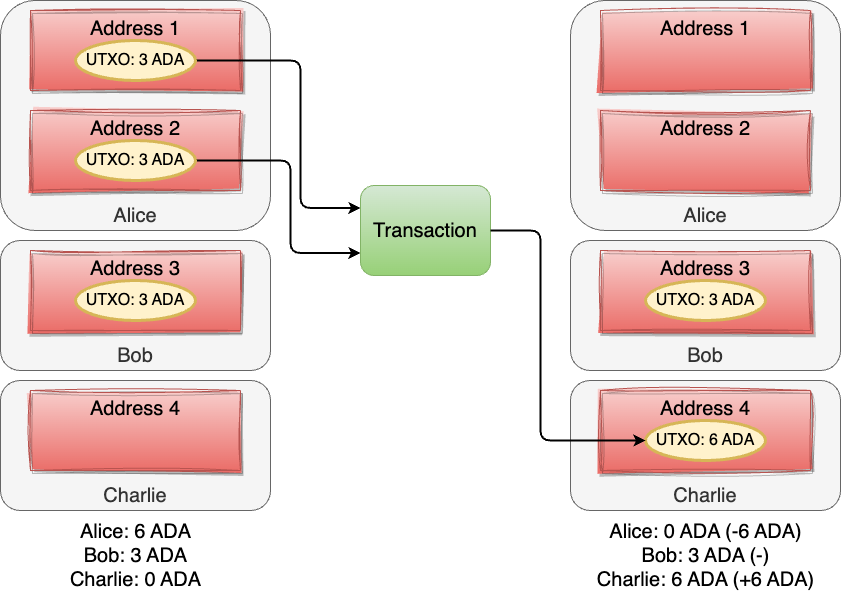}
    \caption{\small Example scenario 2: Value transfer from different addresses of the same entity\\}
    \label{fig:scenario_2a}
\end{figure}

\begin{figure}[h]
    \centering
    \includegraphics[width=0.65\textwidth]{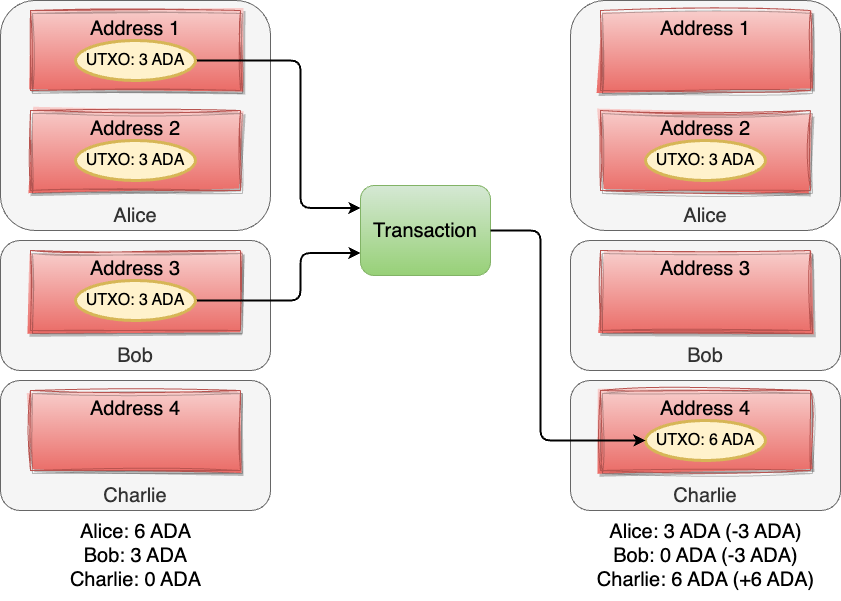}
    \caption{\small Example scenario 2: Value transfer from different addresses of different entities\\}
    \label{fig:scenario_2b}
\end{figure}

Here are two scenarios that can provide insight into the significance of address clustering.
As the first instance, when a certain amount of value is transferred from Address1 to Address2 through a transaction, while both addresses belong to the same entity, neither the distribution of wealth nor the velocity of assets must be affected in analyses by that part of the transaction because it represents no value transfer in the real world. 
Whereas without knowing that these addresses represent the same entity, the results of such analyses would not be accurate. As shown in Fig.~\ref{fig:scenario_1}, the transaction transfers 3 ADA in the network, while the real amount of transferred value among users is only 1 ADA.
Another example would be when two UTXOs stored in two distinct addresses belonging to a single entity are spent. In this case, the corresponding impact on analyses must be completely different from the case when two similar UTXOs owned by two separate entities are consumed. Fig.~\ref{fig:scenario_2a} and Fig.~\ref{fig:scenario_2b} illustrate the difference between these two cases.\\

As can be conceived by the scenarios above, capturing a reliable address clustering has a key role to play in UTXO-based blockchain analytics. 
The effective heuristics for address clustering in UTXO-based blockchains have been discussed in a number of previous studies
\cite{b1,b15,b5,b6,b7,b8,b9,b10,b11}. 
However, since Cardano has extended the UTXO-based accounting model by introducing the Extended Unspent Transaction Output (EUTXO)-based model\cite{b20} to support smart contracts, the already-existing heuristics developed for other blockchains are not directly applicable to Cardano.
In this paper, we focus on Cardano address clustering and propose two heuristics for the EUTXO model.
We apply our heuristics to the Cardano blockchain data to create clusters of addresses, with each cluster representing an entity controlling all addresses within the cluster.

In the following, a brief explanation of the basic concepts and preliminaries of the Cardano blockchain is presented in \textbf{Section~\ref{sec:background}}. 
Afterward, an overview of the literature is provided in \textbf{Section~\ref{sec:related_work}} by referring to relevant work proposed in the blockchain address clustering area.
The approach followed in this work for address clustering is presented in \textbf{Section~\ref{sec:approach}}. The implementation methods, as well as the analytics results, are discussed in \textbf{Section~\ref{sec:results}}. Finally, \textbf{Section~\ref{sec:conclusion}} presents the conclusions.

\section{Background and Preliminaries}\label{sec:background}

\subsection{\textbf{Cardano Consensus Protocol}}
Cardano consensus protocol, Ouroboros, is designed based on Proof of Stake\cite{b18}. 
Each user holding a stake in Cardano can earn passive rewards by participating in the block validation process. 
This participation can be achieved by either 1) establishing a new stake pool and pledging/delegating stake to this pool or 2) delegating stake to an already-existing pool.
In order to delegate their stakes to a pool, users must first register a stake address, and then obtain a delegation certificate for that address.
A stake pool typically consists of three components: 1) one stake pool operator (SPO), 2) one/multiple stake pool owners, and 3) one/multiple stake delegators.
In the Cardano protocol, time is divided into epochs that last 432,000 slots, each lasting one second.
In every time slot, one of the online registered pools is randomly chosen by the protocol to act as the slot leader, the pool eligible to produce the next block. 
Later on, validation rewards are calculated proportional to the amount of stakes pledged or delegated to the pool and are distributed by the protocol among the pool components.
More specifically, to become redeemable and consumable, these rewards must be first withdrawn via transactions that convert them into spendable UTXOs.

\vspace{0.5cm}

\subsection{\textbf{Cardano Accounting Model}}
Cardano basically falls in the category of UTXO-based blockchains. 
Cardano's native cryptocurrency is called ADA.
Having said that, Cardano has introduced the EUTXO-based model, which extends the UTXO-based model to support smart contracts.
A smart contract is a low-level code script that runs synchronously on multiple blockchain nodes\cite{b3}.
In contrast to Bitcoin which does not support complex smart contracts but only simple scripts, Cardano's native programming language for smart contracts, Plutus, is a Turing-complete language written in Haskell and is capable of implementing any logic\cite{b22}.
In Cardano's EUTXO model, each UTXO is locked by a payment address, which refers to 1) a payment key, 2) a simple script (e.g., multi-signature script), or 3) a Plutus smart contract\cite{b23,b24,b25}. 
A crucial point to note here is that the scripts and smart contracts can maintain assets in their own addresses, and these assets can be spent if and only if the conditions pre-defined in the script/contract are satisfied by the spending transaction.

\vspace{0.5cm}

\subsection{\textbf{Cardano Address Types}}
Highlighting the features relevant to our clustering method, Cardano addresses can be categorized into two major categories as follows\cite{b17}.

\begin{itemize}
    \setlength{\itemindent}{-2em}
    \item {\textbf{Payment Addresses}}: Only this type of address can own/spend a UTXO. A payment address can be either a Shelley address or a Byron address.
    \begin{itemize}
        \setlength{\itemindent}{-4em}
        \item {\textbf{Shelley}}: A Shelley address has two main parts.
        
        \begin{itemize}
            \setlength{\itemindent}{-6em}
            \item {\textbf{payment part}}: This part refers to 
            1) a payment key, 
            2) a simple script, or 
            3) a Plutus smart contract that owns funds stored in the Shelley address.
            
            \item {\textbf{delegation part}}: This part may refer to a stake address that owns stake rights of funds stored in the Shelley address.
        \end{itemize}
        
        \item {\textbf{Byron}}: This is an old address format kept in the protocol only for backward compatibility.      
    \end{itemize}

    \item {\textbf{Stake Addresses}}: A stake address owns/controls staking rewards and refers to a stake key. 

\end{itemize}

\section{Related Work}\label{sec:related_work}

There are different blockchain address clustering methods\cite{b1,b29}, such as 
heuristic-based~\cite{b6,b8,b9,b10,b11,b15,b28,b43,b44,b45}, off-chain information-based\cite{b40,b41,b42}, and behavior-based methods\cite{b38,b39}, the last of which includes data mining-based\cite{b30,b31,b32,b33,b34} and deep learning-based methods\cite{b35,b36,b37}.
This section, however, describes only a number of heuristic-based studies that focus on UTXO-based blockchain address clustering and are more relevant to our work.
Apart from these studies, several heuristics have also been explored for clustering and de-anonymizing the addresses in account-based blockchains such as Ethereum\cite{b12,b13} and Ripple\cite{b14}, which are out of the scope of this paper.

First of all, Nakamoto hinted at the possibility that multiple input addresses of a transaction may be associated with the same person or entity \cite{b2}. 
Based on this assumption, Reid and Harrigan\cite{b6} later proposed the multi-input heuristic method as an effective identity detection method. 
In their paper, they also talked about the detection of change addresses as another heuristic; where the change address in a transaction indicates an output address that belongs to the sender and receives excess input.
As part of their study, Androulaki et al.\cite{b8} analyzed Bitcoin's privacy provisions by applying multi-input and change address heuristics to recognize entities in the network. 
Their assumption was that new change addresses were used to receive transactions' changes. 
According to their findings, even when the users used new addresses for each transaction, almost 40\% of them were still identifiable.
Meiklejohn et al.\cite{b9}, assuming that change addresses only have one input, broadened the change heuristic to include transactions with three or more outputs. 
Ortega et al.\cite{b10} proposed a new change address heuristic by choosing the output address whose value had more decimals to be the change address. Their change address heuristic was indeed based on the assumption that real outputs usually have reduced decimals.
Nick\cite{b11} selected the output address whose value was smaller than all inputs of a transaction as the optimal change address.
In that study, it was assumed that crypto wallets typically do not spend unnecessary UTXOs.

Zhang et al.\cite{b43} modified the one-time change address heuristic by discarding the change addresses that were reused as non-change addresses after the change address was made.
On the basis of multi-conditional recognition, Liu et al.\cite{b44} developed a one-time change address identification algorithm to cluster the Bitcoin addresses.
In comparison to other heuristics, they found that their method can reliably detect almost 12\% more one-time change addresses.
He et al.\cite{b45} improved the change address heuristic and proposed a new heuristic based on the different number of output address transactions.
They used six heuristics, including multi-input and change address heuristics, to recognize entities by address clustering. 
They also added conditional constraints to enhance the accuracy of the identified change addresses and hasten the algorithm's convergence.

Combining the heuristics implemented in the BlockSci\cite{b46} C++ library, Campajola et al.\cite{b15} developed their own heuristic-based address clustering algorithms, most of which relied on the detection of change addresses in transactions.
In their study, they employed logical combinations of five heuristics, such as multi-input and change address heuristics, in order to minimize the number of false positives.
Adapting from these five heuristics, Rafati Niya et al.\cite{b26} for the first time, performed an address clustering to identify the addresses belonging to the same wallet in the Cardano blockchain.
They also conducted an analysis of stake balance distribution, reward distribution, and wealth concentration in the Cardano blockchain.

However, according to our knowledge, none of the above-mentioned heuristics has yet been customized in the literature in order to fit the specific features of the EUTXO model used in the Cardano blockchain.

\section{Approach}\label{sec:approach}

In this work, we study address clustering in the Cardano blockchain by applying a heuristic-based approach. Inspired by the heuristics that have already been developed for UTXO-based blockchains, two heuristics derived from the specific features of the EUTXO model are proposed in the following. 
The first heuristic is a modified version of the multi-input heuristic\cite{b6}, which is used for other UTXO-based blockchains like Bitcoin as outlined in \textbf{Section~\ref{sec:related_work}}. The second heuristic, however, is inspired by the staking and delegation mechanisms inherent in the Cardano consensus protocol and is unique to this protocol.

\begin{itemize}

\item {\textbf{Heuristic 1 (modified multi-input heuristic)}}: All Byron payment addresses, and also all Shelley payment addresses with a payment part referring to a payment key, appearing in the inputs list of a single transaction are assumed to belong to the same entity. 
This assumption stems from the fact that regular Cardano wallets, such as Daedalus, Yoroi, and Coinbase, support only generating transactions that spend UTXOs from wallets controlled by a single user. 
Particularly, note that Shelley payment addresses whose payment part refers to a simple script or a smart contract are excluded from this heuristic. 
This is because, when generating a valid transaction, there is no restriction on the selection of input scripts/contracts, i.e., entities can choose any previously created script/contract they wish as input---provided that the conditions specified in the script/contract are fulfilled by the transaction. 
It should also be noted that this heuristic is different from the multi-input heuristic\cite{b6}, which was proposed for address clustering in Bitcoin and considered all input addresses of a transaction to be owned by the same entity.

\item {\textbf{Heuristic 2 (staking heuristic)}}: All Shelley payment addresses whose delegation part refers to the same stake key are assumed to belong to the same entity. 
This assumption arises from the fact that the entity holding the private key associated with a stake key can unconditionally withdraw the staking rewards stored in the corresponding stake address and send them to any desired payment address.
Accordingly, in a normal situation and in the absence of high levels of off-chain trust, it is quite unlikely for one entity to transfer its stake rights to another.

\end{itemize}

Previous studies have also discussed a few other address clustering heuristics for UTXO-based blockchains, including "optimal change," "new address creation," "address reuse," and "peeling chain" heuristics~\cite{b1,b15,b8,b9,b10,b11,b43,b44,b45}. Although these heuristics might be applicable to the Cardano blockchain as well, this paper focuses exclusively on the new heuristics above that are tailored specifically for Cardano.

\section{Implementation and Results}\label{sec:results}

In this section, we present the procedure and results of applying the heuristics proposed in \textbf{Section \ref{sec:approach}} to all the Cardano transactions from September 2017 to January 2023 in order to obtain the clustering result of all addresses that have appeared as inputs or outputs of transactions during this period of time.

\vspace{0.3cm}

\subsection{\textbf{Dataset, Hardware, and Algorithm}}
To cluster the Cardano addresses, first, a Cardano node was set up, which consists of a component called Cardano DBsync. DBSync is the Cardano node's default indexer that collects and stores on-chain data in a PostgreSQL database\cite{b48}. After setting the node up, through joining three tables from the database---i.e., tx, tx\_out, and tx\_in---blockchain's historical transactions data was extracted and stored for further processing.
Afterward, using PySpark library version 3.3.1, a Python script loaded this data into memory, allowing the inputs and outputs to be analyzed efficiently in accordance with the heuristics.
The Python script was run on a Debian 11 machine with 64 AMD EPYC 3.40 GHz CPU cores and 512 GB memory. 
For further improvement of efficiency, the Union-Find algorithm\cite{b4} was adopted to link the payment addresses during the clustering process.

\vspace{0.5cm}

\subsection{\textbf{Clustering Results}}

\begin{figure}[h]
    \centerline
    {\includegraphics[width=0.6\textwidth]{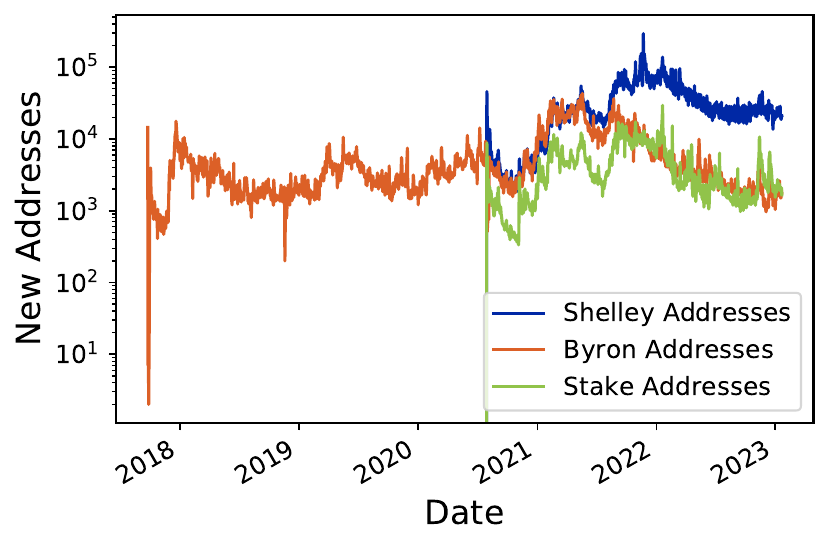}}
    \caption{\small Number of new addresses appeared on the blockchain per day\\}
    \label{fig:new_addresses_per_day}
\end{figure}

Through the history of Cardano until January 2023, the total number of unique payment addresses recorded on the blockchain as owners of UTXOs reaches 40,330,345, including 29,047,961 Shelley addresses and 11,282,384 Byron addresses. 
As well, the total number of stake addresses that have appeared on the blockchain as the delegation part of Shelley addresses is 3,868,049, of which 1,665,652 have been registered. 
This data is summarized in Table~\ref{table:address_types}.
The number of new addresses that have appeared on the blockchain over time is shown in Fig.~\ref{fig:new_addresses_per_day}.

\begin{table}[h]
    \small 
    
    \caption{\small Cardano Address Types}
    \begin{center}
    \begin{tabular}{ c c }

    \hline
    {\textbf{Total Payment Addresses}}&
    40,330,345\\
    
    \hline
    {\textbf{Shelley Addresses}}&
    29,047,961\\

    \hline
    {\textbf{Byron Addresses}}&
     11,282,384\\

    \hline
    {\textbf{Total Stake Addresses}}&
    3,868,049\\

    \hline
    {\textbf{Registered Stake Addresses}}&
    1,665,652\\

    \hline
    
    \end{tabular}
    \label{table:address_types}
    \end{center}
    \vspace{0.3cm}
\end{table}

\begin{figure}[h]
\centerline{\includegraphics[width=0.6\textwidth]{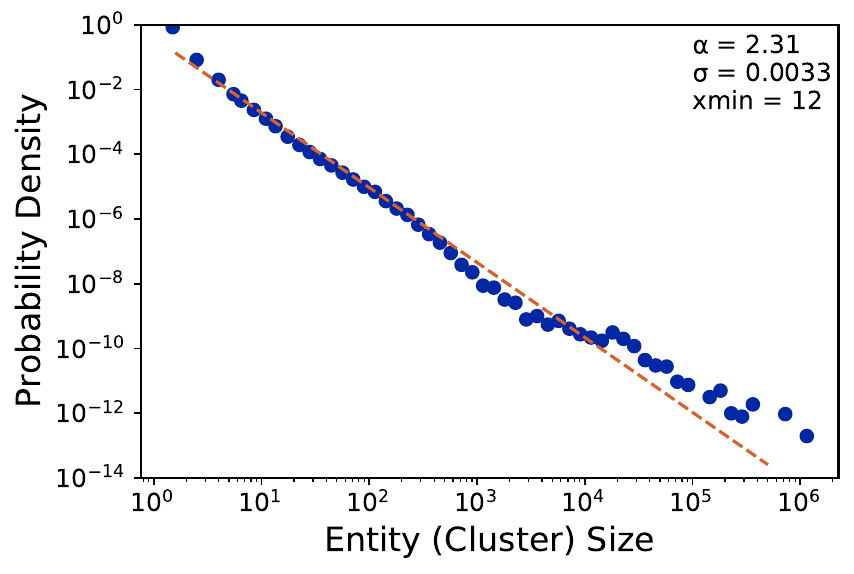}}
    \caption{\small Distribution of addresses per entities (clusters) determined by \textbf{Heuristic 1}\\}
    \label{fig:cluster_mems_heur1}
\end{figure}

We applied \textbf{Heuristic 1} to the payment addresses, which resulted in 19,249,106 distinct clusters. Fig.~\ref{fig:cluster_mems_heur1} illustrates the distribution of addresses in each cluster.

\begin{figure}[h]
    \centering
    \includegraphics[width=0.65\textwidth]{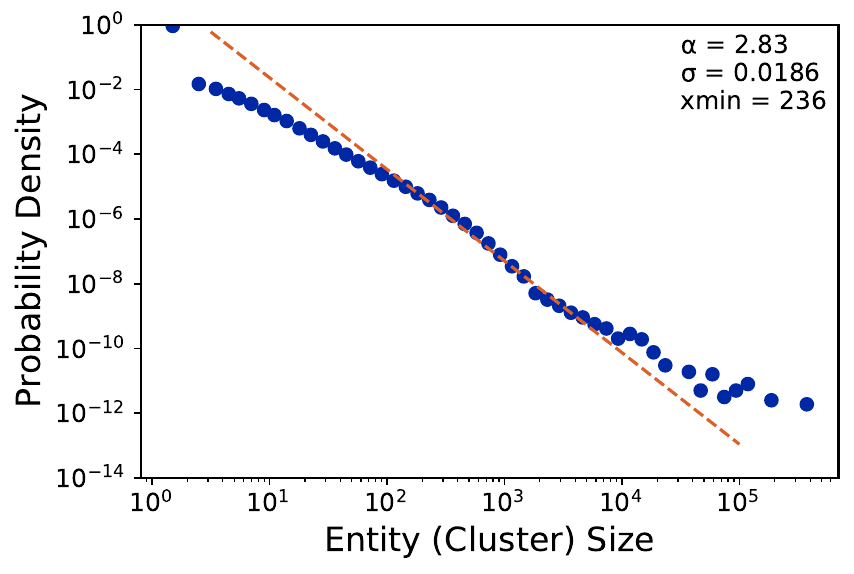}
    \caption{\small Distribution of addresses per entities (clusters) determined by \textbf{Heuristic 2}\\}
    \label{fig:cluster_mems_dist_heur2}
\end{figure}

This clustering indicates that the average number of members (addresses) per cluster is 4.94, excluding 16,310,058 single-member clusters and 349 large clusters that contain more than 1000 members. More notably, there are 9 superclusters with more than 200,000 members.
Likewise, applying \textbf{Heuristic 2}, 1,292,933 non-single-member clusters were detected. The distribution of cluster members according to this analysis is shown in Fig.~\ref{fig:cluster_mems_dist_heur2}.
In this case, after excluding 528 large clusters with at least 1000 members, non-single-member clusters have an average size of 14.63.\\

\begin{figure}[h]
    \centering
    \includegraphics[width=0.6\textwidth]{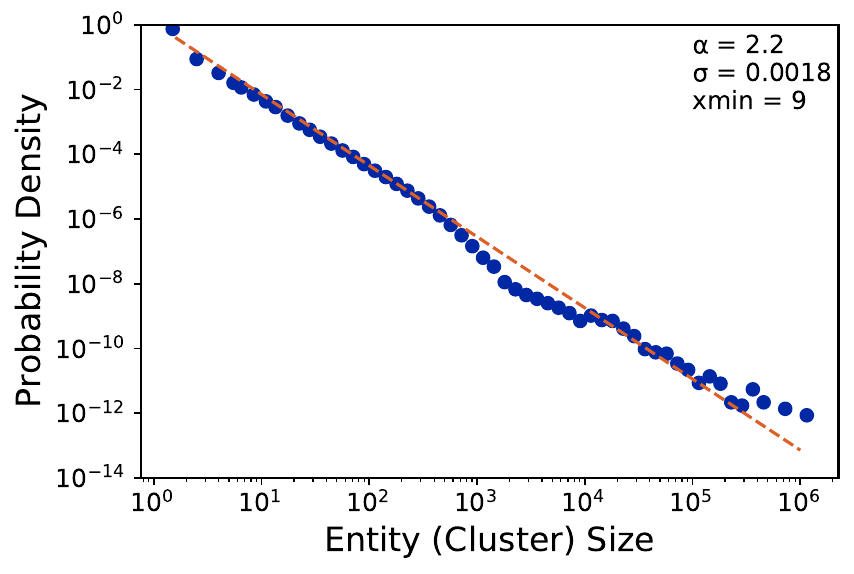}
    \caption{\small Distribution of addresses per entities (clusters) determined by \textbf{Heuristic 1} and \textbf{Heuristic 2}\\}
    \label{fig:cluster_mems_dist_heur1and2}
\end{figure}

In the subsequent step, the clusters obtained by using the two heuristics were merged, providing a more comprehensive address clustering that is illustrated in Fig.~\ref{fig:cluster_mems_dist_heur1and2}.
A summary of clustering results derived from \textbf{Heuristic 1} and \textbf{Heuristic 2} is presented in Table~\ref{table:compare_heur1_heur2}. In particular, the average size of medium-sized entities, that is, clusters with between 2 and 1000 members, is 9.67. 

Our results show that a power law distribution can be fitted to the distribution of entity sizes recognized using \textbf{Heuristic 1} and \textbf{Heuristic 2}. 
A power law probability distribution is in the form of $p(x)\propto~x^{-\alpha}$.
The dashed lines in Fig.~\ref{fig:cluster_mems_heur1}, Fig.~\ref{fig:cluster_mems_dist_heur2}, and Fig.~\ref{fig:cluster_mems_dist_heur1and2} represent a fitted power law distribution obtained by the powerlaw Python package~\cite{b49}. 
The power law distributions have been generated with a fitted parameter $\alpha$ and standard error $\sigma$ for values equal to or more than $xmin$.
It is worth mentioning that combining both heuristics yields a more power-law-like distribution with a smaller standard error and smaller $xmin$.\\

\begin{table}[h]
    \small 
    
    \caption{\small Summary of Clustering Results}
    \begin{center}
    \begin{tabular}{ c c c c }

    \hline
    \textbf{Clustering}&
    \textbf{Heur1}&
    \textbf{Heur2}&
    \textbf{Heur1 and}\\
    
    \textbf{Result}&
    &
    &
    \textbf{Heur2}\\

    \hline
    total \# of clusters&
    19,249,106&
    18,529,342&
    8,805,791\\
    
    average size of clusters$^{\mathrm{a}}$&
    4.94&
    14.63&
    9.67\\

    single-member clusters&
    16,310,058&
    17,236,409&
    6,621,701\\

    large clusters$^{\mathrm{b}}$&
    349&
    528&
    603\\

    superclusters$^{\mathrm{c}}$&
    9&
    3&
    12\\

    \hline
    \multicolumn{4}{l}{$^{\mathrm{a}}${\small excluding large and single-member clusters}}\\
    \multicolumn{4}{l}{$^{\mathrm{b}}${\small clusters with more than 1000 members}} \\
    \multicolumn{4}{l}{$^{\mathrm{c}}${\small clusters with more than 200,000 members}}
    \end{tabular}
    \label{table:compare_heur1_heur2}
    \end{center}
    \vspace{0.3cm}
\end{table}

\begin{figure}[h]
    \centering
    \includegraphics[width=0.6\textwidth]{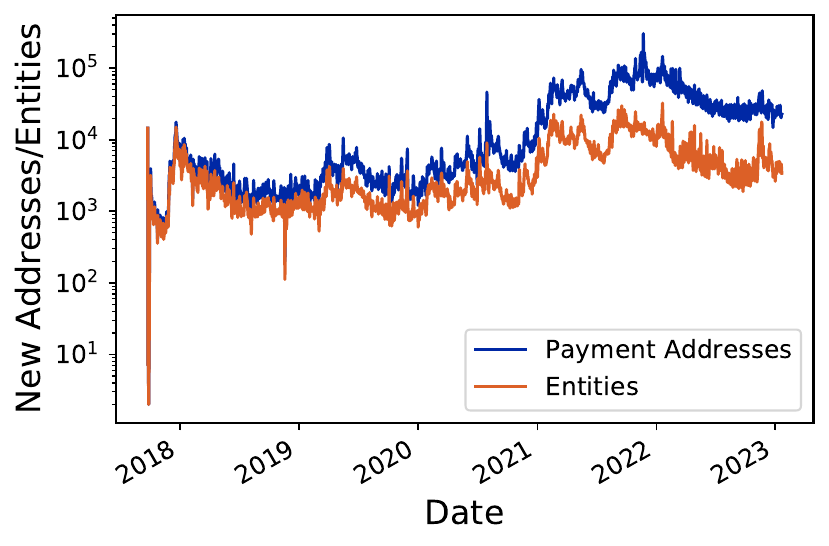}
    \caption{\small Number of new payment addresses and new entities (clusters) determined by \textbf{Heuristic 1} and \textbf{Heuristic 2} appeared on the blockchain per day\\}
    \label{fig:new_entities_per_day}
\end{figure}

Based on this clustering, the number of new entities that have joined the blockchain network over time is calculated and compared with the number of new addresses, the result of which is shown in Fig.~\ref{fig:new_entities_per_day}.

\begin{figure}[h]
    \centering
    \includegraphics[width=0.6\textwidth]{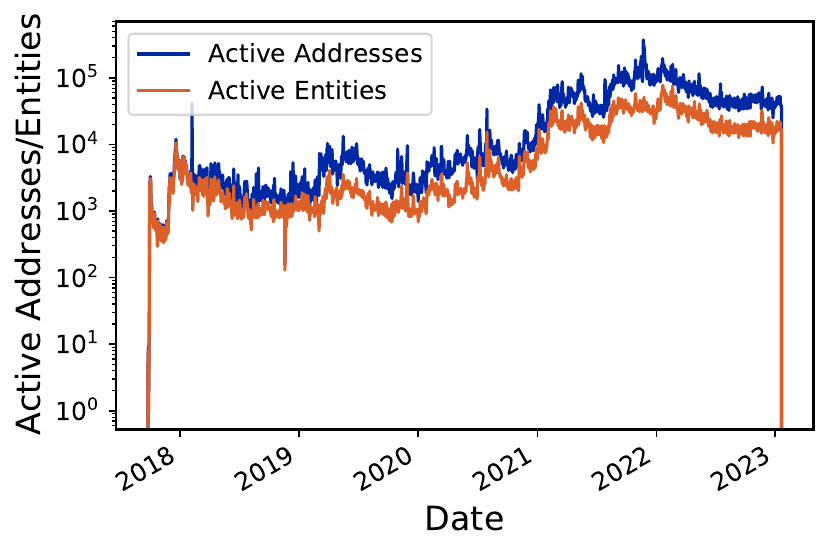}
    \caption{\small Number of active addresses and active entities (clusters) determined by \textbf{Heuristic 1} and \textbf{Heuristic 2} per day}
    \label{fig:active_addresses_per_day}
\end{figure}

In Fig.~\ref{fig:active_addresses_per_day}, the number of active payment addresses per day as well as the number of active entities per day are illustrated. In this figure, if a payment address has appeared in the inputs list of at least one transaction on a given day, that address and consequently its corresponding entity have been considered active.

\begin{figure}[h]
    \centering
    \includegraphics[width=0.6\textwidth]{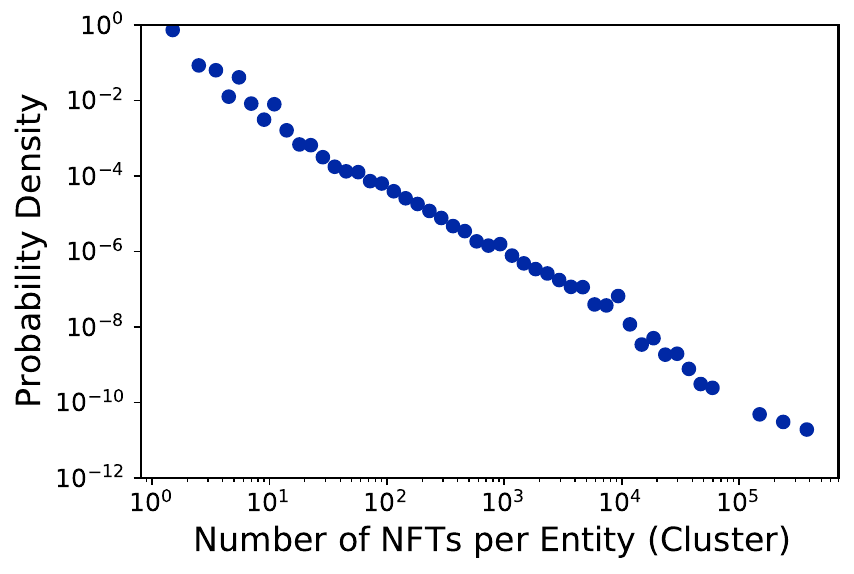}
    \caption{\small Distribution of NFTs minted by entities (clusters) determined by \textbf{Heuristic 1} and \textbf{Heuristic 2} and contributed to NFT minting\\}
    \label{fig:NFTs_per_entity_dist}
\end{figure}

\begin{figure}[h]
    \centering
    \includegraphics[width=0.6\textwidth]{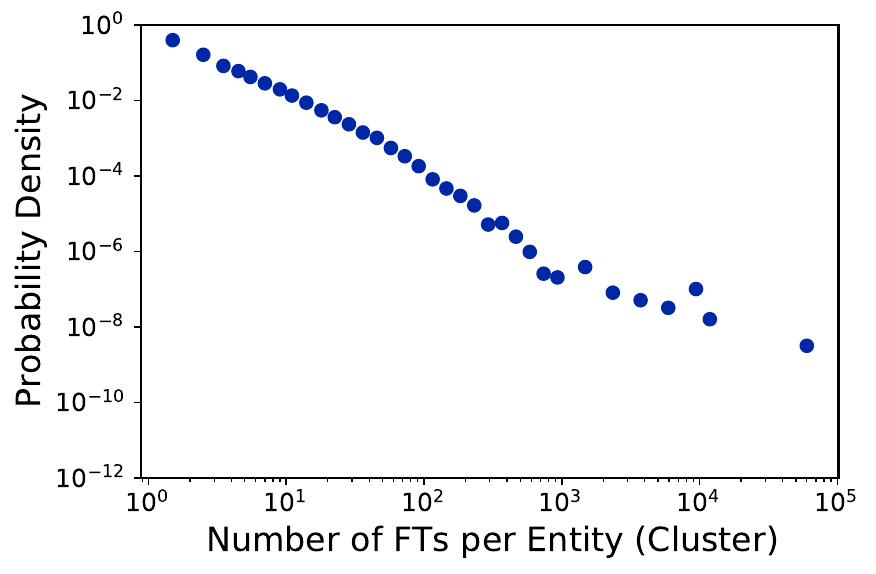}
    \caption{\small Distribution of FTs minted by entities (clusters) determined by \textbf{Heuristic 1} and \textbf{Heuristic 2} and contributed to FT minting\\}
    \label{fig:FTs_per_entity_dist}
\end{figure}

Finally, based on the address clusters indicating the entities in the network, we calculated the distribution of non-fungible tokens (NFTs) and fungible tokens (FTs) minted by each entity, the results of which are presented in Fig.~\ref{fig:NFTs_per_entity_dist} and Fig.~\ref{fig:FTs_per_entity_dist}. According to this analysis, 602,656 entities out of a total of 8,805,791 entities have contributed to NFT minting in the Cardano network. This number equals 22,872 for the entities that have participated in the generation of FTs.

\vspace{0.5cm}

\subsection{\textbf{Discussion and Limitation}}


Although the proposed heuristics generally produce accurate results, they may also produce false positive links: assuming that two addresses are connected together, or owned by the same entity, while in reality, they are not.
For instance, considering \textbf{Heuristic 1}, we have assumed that all UTXOs locked by payment keys listed as inputs of a single transaction are owned by the entity that has generated and signed the transaction via their crypto wallet. 
However, there are some tools, such as "cardano-cli," that are capable of generating complex transactions that carry the signatures of multiple users. 
The input list of such transactions can contain payment keys from different users. 
Consequently, these transactions may insert false positives into our clustering outputs. 
Such false positives are probably responsible for the creation of the superclusters.
As with \textbf{Heuristic 1}, \textbf{Heuristic 2} may also lead to false positives. Indeed, it is technically possible for users to store their assets in Franken addresses, which are Shelley addresses whose delegation part refers to another user's stake address. By doing so, the users can pledge additional stakes to a pool without having to register an additional pool owner on the blockchain, although this requires off-chain trust between the two parties.

In order to improve the accuracy of the proposed address clustering algorithm, these false positive links should be detected and subsequently removed from the output of the algorithm in future work.\\


\vspace{-1em}

\section{Conclusion}\label{sec:conclusion}
Cardano has extended the UTXO-based accounting model by introducing the EUTXO model to support smart contracts. 
Because of this extension, the already-existing heuristics developed for address clustering in other UTXO-based blockchains are not directly applicable to Cardano. 
In this paper, we proposed and implemented two new heuristics based on the specific features of Cardano. 
We also presented the clustering results of Cardano addresses based on the proposed heuristics. 
The results of this paper can be used as a basis for further blockchain analysis in Cardano concerning wealth distribution and asset velocity.
The results show that a power law distribution can be fitted to the distribution of entity sizes recognized using our proposed heuristics.
Nevertheless, the proposed heuristics result in the formation of a few superclusters, indicating that the final clustering results still contain false positives.
In future work, it would be advantageous to improve the accuracy of this heuristic-based address clustering method by detecting and eliminating false positives.


\section*{Acknowledgments}
We acknowledge financial support from the Cardano Foundation.

\section*{Author Contributions}

MCH reviewed the literature, performed the experiments, and wrote the initial draft of the manuscript. SRN collected the data, assisted in structuring the scientific writing of the paper, and proofread the manuscript. CJT conceptualized the study, proposed benchmark models, and proofread the manuscript throughout the writing process.
\bibliographystyle{unsrt}  

\end{document}